\documentclass[aps,prb,preprint,amsmath]{revtex4}

\usepackage{dcolumn}
\usepackage{graphicx}

\renewcommand\vec[1]{{\bf #1}}

\begin{document}

\title{Stark effect in GaN/AlN nanowire heterostructures: Influence of strain relaxation and surface states}

\author{Dulce Camacho Mojica}
\author{Yann-Michel Niquet}
\email{yniquet@cea.fr}
\affiliation{CEA-UJF, INAC, SP2M/L\_Sim, 17 rue des Martyrs, 38054 Grenoble Cedex 9, France}

\date{\today}

\begin{abstract}
We model the quantum confined Stark effect in AlN/GaN/AlN heterostructures grown on top of $[0001]$-oriented GaN nanowires. The pyro- and piezoelectric field are computed in a self-consistent approach, making no assumption about the pinning of the Fermi level, but including an explicit distribution of surface states which can act as a source or trap of carriers. We show that the pyro- and piezoelectric field bends the conduction and valence bands of GaN and AlN and transfers charges from the top surface of the nanowire to an electron gas below the heterostructure. As a consequence, the Fermi level is likely pinned near the valence band of AlN at the top surface. The electron gas and surface charges screen the electric field, thereby reducing the Stark effect. The efficient strain relaxation further weakens the piezoelectric polarization. We compute the electronic properties of the heterostructures with a $sp^3d^5s^*$ tight-binding model, and compare the theoretical predictions with the available experimental data. 
\end{abstract}

\maketitle

\section{Introduction}
\label{sectionIntroduction}

Wide band gap nitride semiconductors are now widely used for light emission in the blue and ultraviolet range.\cite{Nakamura96,Ponce97,Ficher04} Thanks to large band offsets, GaN/AlN heterostructures are also promising candidates for fast telecommunication devices based on intersubband transitions\cite{Nevou07} or for high-temperature single photon emitters.\cite{Kao06} One of the specifics of nitride heterostructures is the existence of large internal electric fields due to spontaneous polarization and strains (piezoelectricity).\cite{Bernardini97} These built-in fields might transfer charges in the devices, leading for example to the formation of two dimensional (2D) electron gases at the interfaces between GaN and (Ga)AlN layers.\cite{Ambacher99,Ibbetson00,Jang02,Jogai03,Koley05} They might also separate the electrons from the holes in GaN quantum wells and Stranski-Krastanov (SK) quantum dots, thereby reducing the band gap and oscillator strength (quantum confined Stark effect).\cite{Widmann98,Simon03,Adelmann03,Bretagnon06} It is therefore essential to understand and tailor the electric field in nitride heterostructures to suit a particular application.

Whereas 2D layers usually feature a large density of dislocations, nitride nanowires offer the opportunity to make defect-free heterostructures thanks to the efficient strain relaxation.\cite{Ertekin05,Glas06} Single GaN quantum disks (QDs) between two AlN barriers have for example been grown on top of GaN pillars (20--50 nm diameter) with plasma-assisted molecular beam epitaxy (see Fig. \ref{figsystem}).\cite{Renard08} The exciton and biexciton luminescence of 1 nm thick GaN QDs has been observed, showing the potential of such heterostructures for nitride optoelectronic devices.\cite{Renard08} A strong red shift (below the bulk GaN band gap) has been subsequently observed for larger disk thickness, a signature of the quantum confined Stark effect.\cite{Renard09} The apparent electric field is however smaller than expected from a comparison with GaN/AlN quantum wells.\cite{Adelmann03} The effects of strain relaxation (decrease of the piezoelectricity) and finite lateral size on the electric field,\cite{Wu09} as well as the screening mechanisms are still unclear.

Nitride SK dots have been modeled before with $\vec{k}\cdot\vec{p}$ or tight-binding approaches.\cite{Wu09,Andreev00,Ranjan03,Ristic05,Marquardt08,Schulz09,Williams09} The key role played by the charges transferred by band bending, which screen the electric field, has been emphasized in 2D GaN/GaAlN layers.\cite{Ambacher99,Ibbetson00,Jang02,Jogai03,Koley05} In this paper, we model the electronic properties of $[0001]$-oriented GaN/AlN nanowire heterostructures in an atomistic tight-binding framework.\cite{Niquet06b,Niquet08} We compute strains with a valence force field method,\cite{Camacho09} then the pyro- and piezoelectric field. We account for band bending with a semi-classical Debye-H\"uckel approach, making no assumption about the pinning of the Fermi level, but including a distribution of surface states which act as a source or trap of carriers. We show {\it i}) that the piezoelectric component of the field can be significantly reduced by the efficient strain relaxation in the nanowire geometry ; {\it ii}) that the spontaneous and piezoelectric polarizations create an electron gas at the lower GaN/AlN interface and are likely large enough to create a hole gas in the upper AlN barrier. These electron and hole gases screen the electric field in the GaN QD and reduce the Stark shift ; {\it iii}) that for carefully chosen dot and barrier thicknesses (realized experimentally) the GaN QD is empty at equilibrium, consistent with the observation of exciton and biexciton transitions. We discuss the magnitude of the electric field and the electronic stucture of the QDs as a function of the dimensions of the heterostructure.

The paper is organized as follows: We review the methods in section \ref{sectionMethodology}, then discuss the electric field in GaN/AlN nanowire heterostructures in section \ref{sectionEfield}. We introduce a simple 1D model for the pyro- and piezoelectric field that reproduces the main trends. We finally discuss the electronic structure of the GaN QDs and compare our calculations with the available experimental data in section \ref{sectionelectronic}. We analyze the dependence of the electronic and optical properties of the QDs on the geometry of the heterostructures.

\section{Methodology}
\label{sectionMethodology}

\begin{figure}
\includegraphics[width=.66\columnwidth]{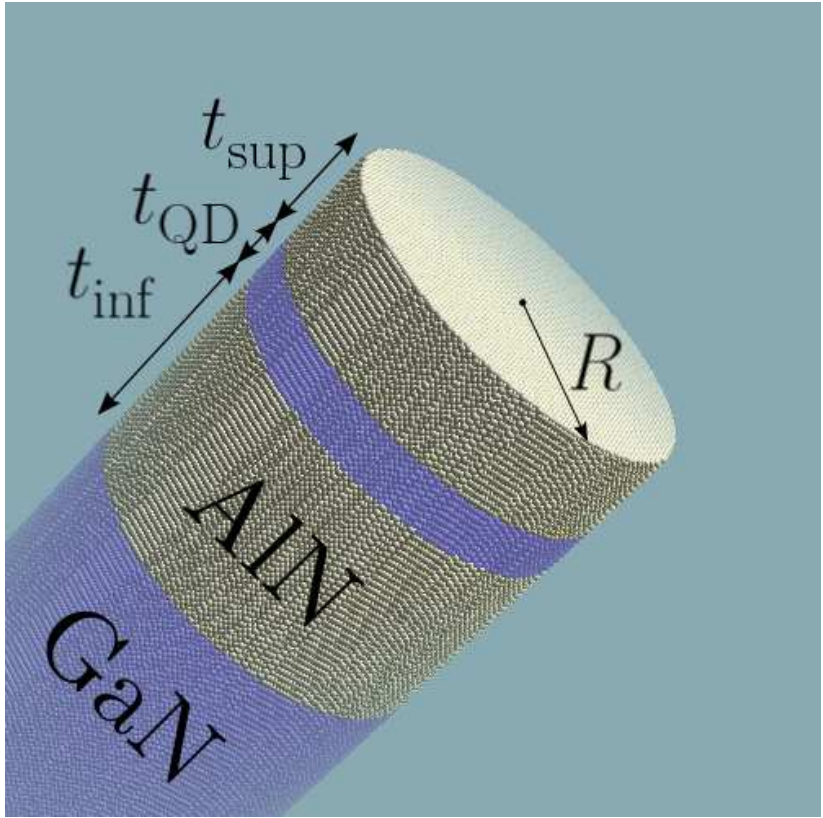}
\caption{(Color online) Structure of the GaN/AlN nanowire heterostructures. The radius of the nanowires is $R=15$ nm.}
\label{figsystem}
\end{figure}

In this section, we introduce the methods used to compute the structural and electronic properties of the GaN/AlN heterostructures.

Each nanowire is modeled as a 30 nm diameter and 150 nm long\cite{notelong} cylindrical GaN pillar oriented along $[0001]$, with the heterostructure on top (we assume metal-face polarity -- the case of N-face polarity will be briefly discussed at the end of paragraph \ref{sectionelectronic}). The heterostructure consists of a lower AlN barrier with thickness $t_{\rm inf}$, a GaN quantum disk with thickness $t_{\rm QD}$ and a upper AlN barrier with thickness $t_{\rm sup}$ (see Fig. \ref{figsystem}). The dangling bonds at the surface of the nanowire are saturated with hydrogen atoms.

The strains in the nanowire are computed with Keating's valence field model.\cite{Keating66} This model provides an atomistic description of the elasticity of tetrahedrally bonded semiconductors. It was originally designed for zinc-blende and ``ideal'' wurtzite materials with equal bond lengths and angles,\cite{Martin72} and has been recently adapted to arbitrary wurtzite materials such as GaN and AlN.\cite{Camacho09} The elastic energy of the nanowire is minimized with respect to the atomic positions using a conjugate gradients algorithm. The strains $\varepsilon_{\alpha\beta}$ on each atom are then calculated from the atomic positions using a method similar\cite{noteeij} to Ref. \onlinecite{Pryor98}.

\begin{table}
\begin{tabular}{lrr}
\toprule
 & GaN & AlN \\
\hline
$P_0$ (C/m$^2$)\footnotemark[1] & $-0.034$ & $-0.090$ \\
$e_{13}$ (C/m$^2$)\footnotemark[1] &  $-0.53$ &  $-0.54$ \\
$e_{33}$ (C/m$^2$)\footnotemark[1] &   $0.89$ &   $1.56$ \\
$e_{15}$ (C/m$^2$)\footnotemark[1]\footnotemark[2] &  $-0.33$ &  $-0.42$ \\
$N_v$ (cm$^{-3}$)\footnotemark[3] & $4.6\times 10^{19}$ & $4.8\times 10^{20}$ \\
$N_c$ (cm$^{-3}$)\footnotemark[3] & $2.3\times 10^{18}$ & $6.3\times 10^{18}$ \\
$E_v$ (eV)\footnotemark[1] &  $0.0$ &   $-0.8$ \\
$E_c$ (eV)\footnotemark[1] & $3.50$ &   $5.45$ \\
$E_b$ (meV)\footnotemark[4] &   $30$ &  $170$ \\
$E_1^+$ (eV) & $0.25$ & $1.00$ \\
$E_2^+$ (eV) & $1.25$ & $3.00$ \\
$E_1^-$ (eV) & $2.25$ & $4.25$ \\
$E_2^-$ (eV) & $3.25$ & $5.25$ \\
\botrule
\end{tabular}
\footnotetext[1]{Ref.~\onlinecite{Vurgaftman03}}
\footnotetext[2]{Ref.~\onlinecite{Schulz09}} 
\footnotetext[3]{Ref.~\onlinecite{Morkoc08}} 
\footnotetext[4]{Ref.~\onlinecite{Tansley92}}
\caption{The material parameters for GaN and AlN.\cite{Vurgaftman03,Schulz09,Morkoc08,Tansley92}}
\label{tabparams}
\end{table}

The pyro- and piezoelectric polarization density is next computed from the strains on each cation (Ga or Al) :
\begin{equation}
\vec{P}=P_0\vec{z}+
\begin{pmatrix}
2e_{15}\varepsilon_{xz} \\
2e_{15}\varepsilon_{yz} \\
e_{31}(\varepsilon_{xx}+\varepsilon_{yy})+e_{33}\varepsilon_{zz}
\end{pmatrix}\,,
\end{equation}
where $z\equiv[0001]$, $P_0$ is the spontaneous polarization, and $e_{13}$, $e_{33}$ and $e_{15}$ are the piezoelectric constants of either GaN or AlN (see Table \ref{tabparams}). Poisson's equation for the pyro- and piezoelectric potential $V_p(\vec{r})$:
\begin{equation}
\kappa_0\vec{\nabla}_\vec{r}\cdot\kappa(\vec{r})\vec{\nabla}_\vec{r}V_p(\vec{r})=\vec{\nabla}_\vec{r}\cdot\vec{P}(\vec{r})
\end{equation}
is then solved on a finite difference grid\cite{noteV} (see appendix I of Ref. \onlinecite{Niquet06b} for details). $\kappa$ is the dielectric constant ($\kappa=9$ inside the nanowire and $\kappa=1$ outside).

The large pyro- and piezoelectric field $\vec{E}_p=-\vec{\nabla}V_p$ in the heterostructure bends the conduction and valence bands and can therefore transfer charges from one part of the system to an other. It is for example known that the spontaneous polarization in GaAlN layers grown on GaN pulls out electrons from the GaAlN surface, which accumulate in a 2D electron gas at the GaN/GaAlN interface.\cite{Ambacher99,Ibbetson00,Jang02,Jogai03,Koley05} These electrons leave positive charges at the GaAlN surface, which can be ionized surface donors, emptied surface states, or even a hole gas. This redistribution of charges creates, in turn, an electric field opposite to $\vec{E}_p$, which can screen the latter to a large extent.

The effects of band bending have been self-consistently computed in a semi-classical Debye-H\"uckel approximation. The local density of electrons,  $n(\vec{r})$, and the local density of holes, $p(\vec{r})$ are calculated as:\cite{Sze}
\begin{subequations}
\label{eqnp}
\begin{eqnarray}
n(\vec{r})&=&N_c F_{1/2}\left[-\beta\left(E_c-eV(\vec{r})-\mu\right)\right] \\
p(\vec{r})&=&N_v F_{1/2}\left[+\beta\left(E_v-eV(\vec{r})-\mu\right)\right] \,,
\end{eqnarray}
\end{subequations}
where $N_c$ and $N_v$ are the effective conduction and valence band density of states of the material at point $\vec{r}$, $E_c$ and $E_v$ are its conduction and valence band edge energies (see Table \ref{tabparams}), $V(\vec{r})$ is the total electrostatic potential, and $\mu$ is the chemical potential or Fermi energy. $F_{1/2}$ is the Fermi integral of order one-half and $\beta=1/(kT)$, where $T=300$ K is the temperature. We have, additionally, assumed that the nanowires were non-intentionally $n$-doped, with a concentration of donor impurities (silicon, oxygen or vacancies) $N_d=2\times10^{17}$ cm$^{-3}$. The density of ionized impurities is:\cite{Sze}
\begin{equation}
N_d^+(\vec{r})=\frac{N_d}{1+2e^{-\beta\left[E_c-E_b-eV(\vec{r})-\mu\right]}}\,,
\label{eqnd}
\end{equation}
where $E_b$ is the binding energy of the donor, which typically ranges from a few tens to a few hundreds of meV.

As mentioned previously, surface states can play an important role in the electrostatics of nitride nanowires. They might act as a source\cite{Ibbetson00,Jang02,Jogai03,Koley05} or as a trap\cite{Schmidt07,Simpkins08} of carriers, effectively pinning the chemical potential in the band gap. Little is however known about the electronic structure of nitride surfaces.\cite{Feenstra05} On one hand, density functional theory (DFT) calculations on reconstructed GaN and AlN surfaces\cite{Northrup97,Fritsch98,Segev06,Walle07,Miao09} suggest the existence of occupied (donor-like) surface states above the valence band edge and empty (acceptor-like) surface states below the conduction band edge (as expected from simple considerations). On the other hand, the extensive literature about 2D electron gases in $[0001]$ GaN/GaAlN heterostructures\cite{Ibbetson00,Jang02,Jogai03,Koley05} suggests the existence of dense ($\simeq 10^{13}$ cm$^{-2}$eV$^{-1}$) surface donor states only $\simeq 1.5$ eV below the conduction band of Ga$_{1-x}$Al$_{x}$N alloys ($x\simeq0.4$). Although the nature of these surface donors is still debated, oxygen has often been put forward.\cite{Dong06} It is not clear however that the same picture holds for non-polar Ga(Al)N surfaces and for surfaces of {\it pure} AlN, where the oxide is not the same. The situation is particularly tricky in nanowires, which expose different (polar and non-polar) surfaces. For the sake of simplicity, we assume in this work the existence of a uniform density of occupied surface states in the $[E_1^+, E_2^+]$ energy range above the valence band edge, and of a uniform density of empty surface states in the $[E_1^-, E_2^-]$ energy range below the conduction band edge. The density of ionized occupied surface states is therefore:
\begin{equation}
N_s^+(\vec{r})=kT\,D_s^+\ln\frac{1+\frac{1}{2}e^{\beta\left[E_2^+-eV(\vec{r})-\mu\right]}}{1+\frac{1}{2}e^{\beta\left[E_1^+-eV(\vec{r})-\mu\right]}}\,,
\label{eqnsp}
\end{equation}
while the density of electrons trapped in the empty surface states is:
\begin{equation}
N_s^-(\vec{r})=kT\,D_s^-\ln\frac{1+2e^{-\beta\left[E_1^--eV(\vec{r})-\mu\right]}}{1+2e^{-\beta\left[E_2^--eV(\vec{r})-\mu\right]}}\,.
\label{eqnsm}
\end{equation}
$D_s^+$ and $D_s^-$ are the density of occupied and empty surface states, respectively (per unit surface and energy). The values of $E_1^+$, $E_2^+$, $E_1^-$ and $E_2^-$ used in this work are also reported in Table \ref{tabparams}. They are representative of {\it ab initio} calculations,\cite{Northrup97,Fritsch98,Segev06,Walle07,Miao09} and reproduce the pinning of the Fermi level on non-polar GaN surfaces.\cite{Simpkins08} We have varied $D_s^+=D_s^-$ between $5\times10^{12}$ cm$^{-2}$eV$^{-1}$ and $5\times 10^{13}$ cm$^{-2}$eV$^{-1}$. Their effects will be discussed in paragraph \ref{sectionEfield}. We will show, in particular, that the electric field in the QD is weakly dependent on the model for the surface states up to large $D_s^+$ and $D_s^-$.

In practice, the carrier densities $n(\vec{r})$ and $p(\vec{r})$ are computed on each Ga, Al and N atom, while the surface state densities $N_s^+(\vec{r})$ and $N_s^-(\vec{r})$ are computed on each hydrogen atom. The charge on each atom is then transferred to the finite difference mesh, and Poisson's equation for the total electrostatic potential $V(\vec{r})$ is solved self-consistently with the Newton-Raphson method:\cite{numrecipes}
\begin{eqnarray}
&&\kappa_0\vec{\nabla}_\vec{r}\cdot\kappa(\vec{r})\vec{\nabla}_\vec{r}V(\vec{r})=\vec{\nabla}_\vec{r}\cdot\vec{P}(\vec{r}) \nonumber \\
&+&\left[n(\vec{r})-p(\vec{r})-N_d^+(\vec{r})+N_s^-(\vec{r})-N_s^+(\vec{r})\right]e\,.
\label{eqv}
\end{eqnarray}
The chemical potential $\mu$ is adjusted to ensure overall charge neutrality of the nanowire.

Finally, the electronic structure of the GaN QD in the potential $V(\vec{r})$ is computed with a $sp^3d^5s^*$ tight-binding model.\cite{Slater54,DiCarlo03,Delerue05,Jancu02} In order to access the relevant states directly, a slice containing the GaN QD and 4 nm of each AlN barrier is cut from the nanowire. The bonds broken by this operation are saturated with hydrogen atoms, and a few conduction and valence band states are computed with a Jacobi-Davidson algorithm.\cite{Sleijpen00,Niquet06a} The convergence of the electronic structure of the QD with respect to the thickness of the AlN barriers has been checked. The above methodology has been implemented in an in-house code called {\tt TB\_Sim}.

\section{The electric field in GaN/AlN nanowire heterostructures}
\label{sectionEfield}

\begin{figure}
\includegraphics[width=.66\columnwidth]{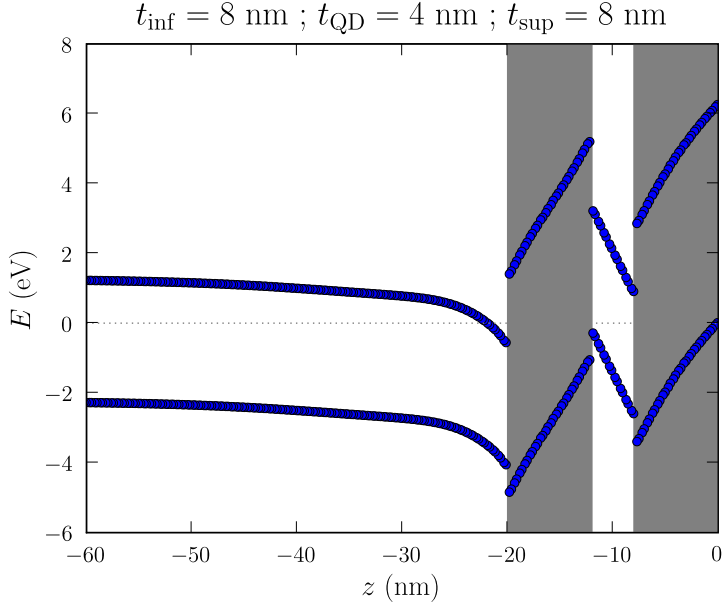}
\caption{(Color online) The conduction band edge energy $\varepsilon_c(\vec{r})=E_c-eV(\vec{r})$ and the valence band edge energy $\varepsilon_v(\vec{r})=E_v-eV(\vec{r})$ along the axis of a nanowire with $t_{\rm QD}=4$ nm and $t_{\rm inf}=t_{\rm sup}=8$ nm. The reference of energy is the chemical potential $\mu=0$. The position of the AlN barriers is outlined in gray, and the top surface is at $z=0$.}
\label{figbands_8_8}
\end{figure}

In this section, we discuss the electric field in GaN/AlN nanowire heterostructures. We first analyze a particular case as an example. We then show that the electric field can be reproduced by a simple 1D model in a wide range of dimensions. We finally discuss the main trends as a function of the geometry of the heterostructures.

\subsection{Example}

We focus as an illustration on a 30 nm diameter nanowire with a $t_{\rm QD}=4$ nm thick GaN QD and $t_{\rm inf}=t_{\rm sup}=8$ nm thick AlN barriers. We set $D_s^+=D_s^-=10^{13}$ cm$^{-2}$eV$^{-1}$.

The conduction band edge energy $\varepsilon_c(\vec{r})=E_c-eV(\vec{r})$ and the valence band edge energy $\varepsilon_v(\vec{r})=E_v-eV(\vec{r})$ are plotted along the axis of the nanowire in Fig. \ref{figbands_8_8}. The reference of energy for this plot is the chemical potential $\mu=0$. The top of the nanowire is located at $z=0$ and the position of the AlN barriers is outlined in gray.

The band discontinuities at the GaN/AlN interfaces are clearly visible. The heterostructure undergoes a strong vertical electric field, which is almost homogeneous in the AlN barriers and in the GaN QD. The latter is empty of carriers (electrons and holes). The chemical potential is however pinned at the valence band edge at the top surface of the nanowire, and crosses the conduction band at the interface with the GaN pillar. Electrons therefore accumulate in the GaN pillar, while holes accumulate in the upper AlN barrier.

\begin{figure}
\includegraphics[width=.66\columnwidth]{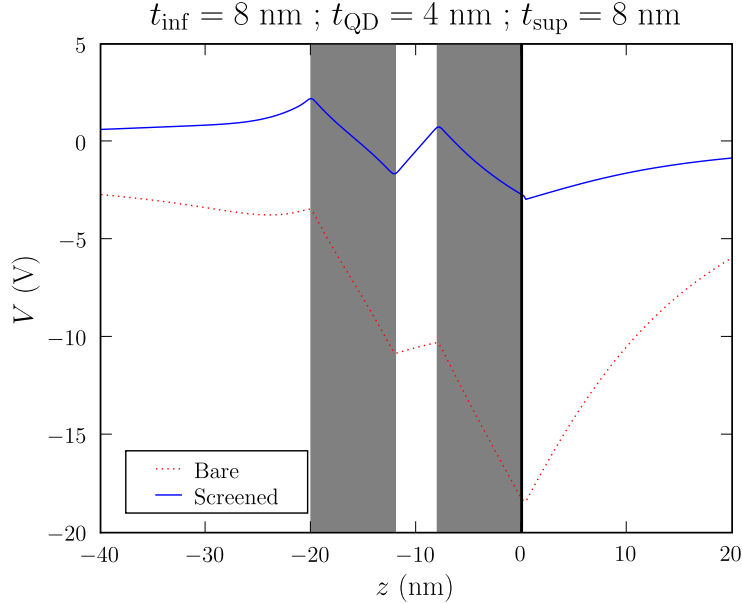}
\caption{(Color online) The bare (unscreened) and screened pyro- and piezoelectric potentials along the axis of the nanowire ($t_{\rm QD}=4$ nm, $t_{\rm inf}=t_{\rm sup}=8$ nm). The position of the AlN barriers is outlined in gray, and the top surface is at $z=0$.}
\label{figvp_8_8}
\end{figure}

This redistribution of charges follows from the pyro- and piezoelectric polarizations. Leaving aside piezoelectricity for the moment, the spontaneous polarization in GaN is $P_z=-0.034$ C/m$^2$, while the spontaneous polarization in AlN is $P_z=-0.090$ C/m$^2$. This polarization is equivalent to a distribution of charges $\sigma=-0.090$ C/m$^2$ at the top surface, and $\sigma=\pm(0.090-0.034)=\pm0.056$ C/m$^2$ at each GaN/AlN interface. Such a charge distribution, if unscreened, would create huge vertical electric fields and potentials of the order of 10 to 20 V in the nanowire (see Fig. \ref{figvp_8_8}).

The pyro- and piezoelectric field however bends the conduction and valence bands and tends to draw positive charges at the top of the nanowire, which screen the polarization. The potential actually rises the occupied surface states of the upper AlN barrier above the Fermi energy. They therefore empty, leaving positive charges at the surface and releasing electrons in the GaN pillar. At moderate electric field, the surface states would be able to provide enough charge to reach equilibrium, and the chemical potential would lie in the band gap at the top AlN surface. Here the electric field is however large enough to empty the $N_{\rm tot}^+=D_s^+(E_2^+-E_1^+)=2\times10^{13}$ cm$^{-2}$ occupied surface states. The chemical potential then sinks into the valence band; a gas of holes forms at the top surface and provides the missing charges. 

\begin{figure}
\includegraphics[width=.66\columnwidth]{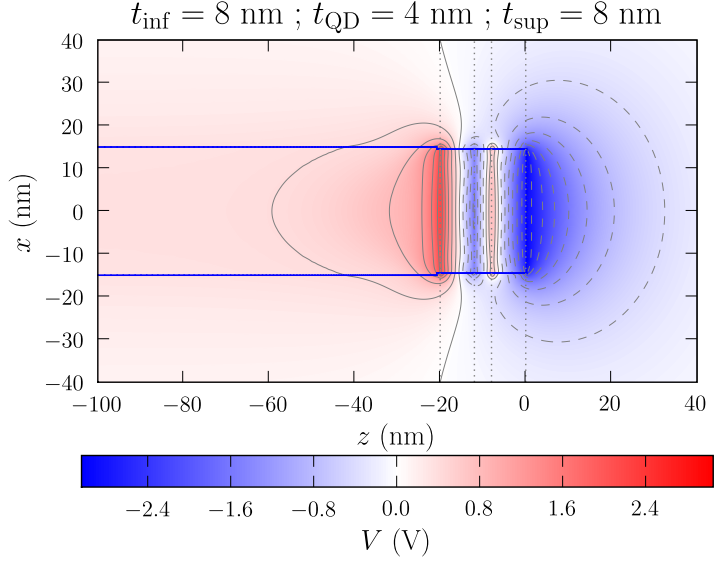}
\caption{(Color online) The electrostatic potential $V(\vec{r})$ in the $(xz)$ plane containing the axis of the nanowire ($t_{\rm QD}=4$ nm, $t_{\rm inf}=t_{\rm sup}=8$ nm). The GaN and AlN layers are delimited by dotted lines.}
\label{figV_8_8}
\end{figure}

According to this picture, the charge in the system is mostly distributed at the top AlN surface and at the GaN/AlN interfaces. As a consequence, the electric field is typical of a series of parallel plate capacitors, being almost homogeneous in the GaN QD and AlN barriers. This is further emphasized in Fig. \ref{figV_8_8}, which represents the electrostatic potential $V(\vec{r})$ in a $(xz)$ plane containing the axis of the nanowire. The equipotential lines are indeed parallel to the interfaces. The electrostatic corrections due to the finite cross section of the nanowire are therefore limited in the GaN QD and barriers in this range of dimensions.

The effective density of states in the conduction and valence bands of GaN and AlN are large enough to ``lock'' the potential at the interface with the GaN pillar and at the top surface (as a small variation of potential leads to exponential variations of the charge densities once the Fermi energy is in the bands). Hence,
\begin{subequations}
\label{eqmu}
\begin{equation}
\mu\simeq E_v({\rm AlN})-eV(z=0)
\end{equation}
at the top surface, and
\begin{equation}
\mu\simeq E_c({\rm GaN})-eV(z=-t_{\rm het})
\end{equation}
\end{subequations}
at the interface $z=-t_{\rm het}=-(t_{\rm inf}+t_{\rm QD}+t_{\rm sup})$ with the GaN pillar. The voltage drop $\Delta V=V(z=-t_{\rm het})-V(z=0)$ across the heterostructure is therefore:
\begin{eqnarray}
e\Delta V&\simeq& E_c({\rm GaN})-E_v({\rm AlN}) \nonumber \\
&\simeq&E_g({\rm GaN})+E_v({\rm GaN})-E_v({\rm AlN})\,. \label{eqDV}
\end{eqnarray}
The voltage drop across the heterostructure is thus primarily defined by the band gap $E_g({\rm GaN})$ of GaN and the valence band offset between GaN and AlN once the Fermi energy is pinned in the valence band of AlN at the top surface. The validity of this assumption will be discussed in the next paragraphs.

\begin{figure}
\includegraphics[width=.66\columnwidth]{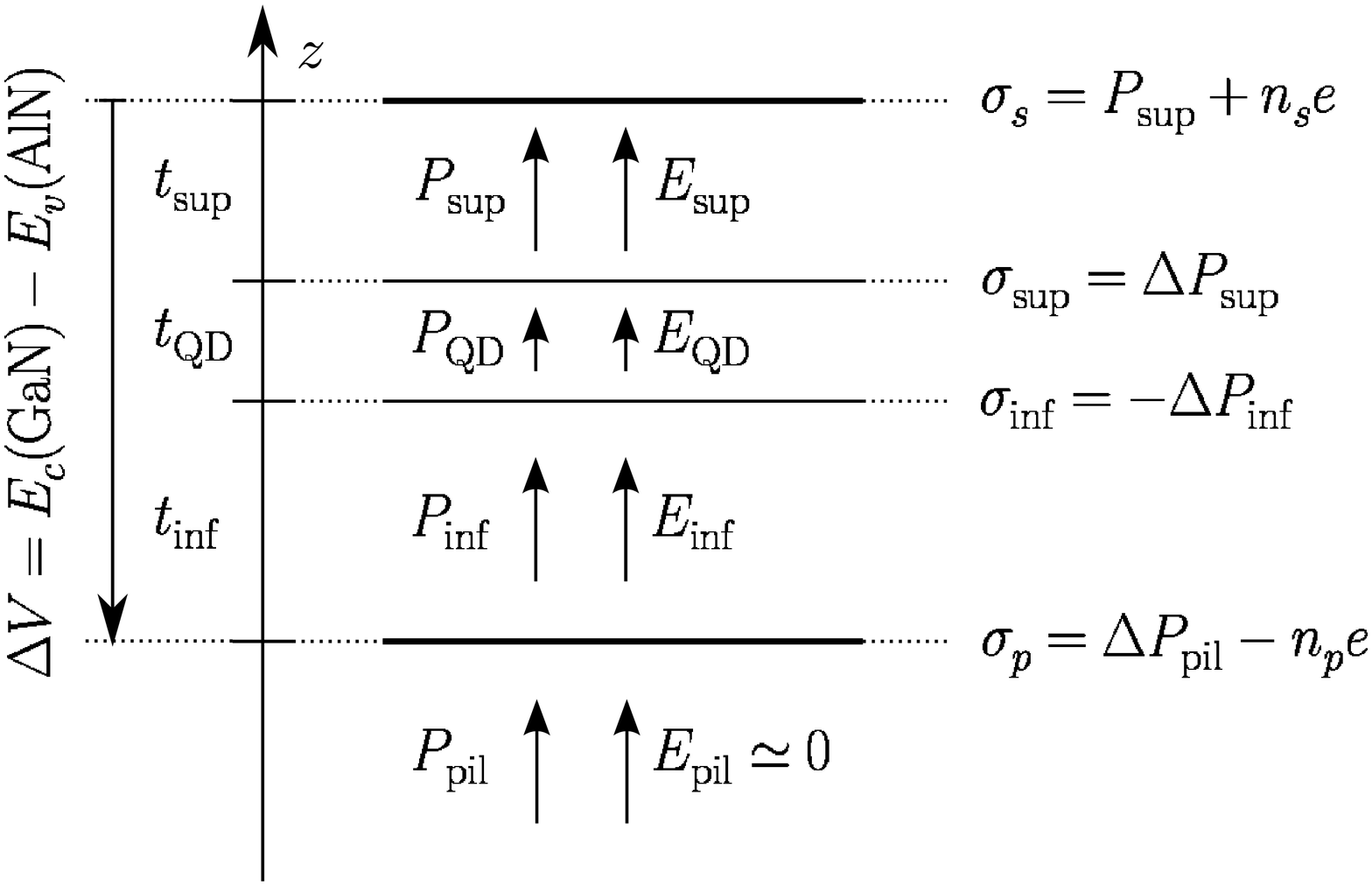}
\caption{The 1D model used for the analysis of the electric field in the GaN QD.}
\label{figModel}
\end{figure}

\subsection{A simple 1D model}

We can derive a simple 1D model for the electric field $E_{\rm QD}$ in the GaN QD from the above observations. For that purpose, we neglect finite size effects ($R\to\infty$) and doping. We assume that the polarization is homogeneous in the lower AlN barrier ($P_z=P_{\rm inf}$), GaN quantum disk ($P_z=P_{\rm QD}$) and upper AlN barrier ($P_z=P_{\rm sup}$, see Fig. \ref{figModel}). This polarization is equivalent to a charge density $\sigma_{\rm sup}=\Delta P_{\rm sup}=P_{\rm QD}-P_{\rm sup}$ on the upper QD interface and $\sigma_{\rm inf}=-\Delta P_{\rm inf}=-(P_{\rm QD}-P_{\rm inf})$ on the lower QD interface. We also assume that the difference of potential $\Delta V$ across the heterostructure is set by band structure effects [Eq. (\ref{eqDV}) or equivalent if other pinning of the Fermi level]. The electric field is then homogeneous in each layer and fulfill the continuity and integral equations:
\begin{subequations}
\begin{eqnarray}
&&\kappa_0\kappa\left(E_{\rm inf}-E_{\rm QD}\right)=\Delta P_{\rm inf} \\
&&\kappa_0\kappa\left(E_{\rm sup}-E_{\rm QD}\right)=\Delta P_{\rm sup} \\
&&t_{\rm inf}E_{\rm inf}+t_{\rm QD}E_{\rm QD}+t_{\rm sup}E_{\rm sup}=\Delta V\,.
\end{eqnarray}
\end{subequations}
We therefore get:
\begin{equation}
E_{\rm QD}=-\frac{1}{\kappa_0\kappa}\frac{t_{\rm inf}+t_{\rm sup}}{t_{\rm het}}\Delta\bar{P}+\frac{\Delta V}{t_{\rm het}}
\label{EqEQD}
\end{equation}
where:
\begin{equation}
\Delta\bar{P}=\frac{t_{\rm inf}\Delta P_{\rm inf}+t_{\rm sup}\Delta P_{\rm sup}}{t_{\rm inf}+t_{\rm sup}}
\end{equation}
is an average polarization discontinuity at the interfaces of the QD. Additionally, the electric field in the barriers is:
\begin{subequations}
\label{eqEinfsup}
\begin{eqnarray}
E_{\rm inf}&=&\frac{1}{\kappa_0\kappa}\left[\Delta P_{\rm inf}-\frac{t_{\rm inf}+t_{\rm sup}}{t_{\rm het}}\Delta\bar{P}\right]+\frac{\Delta V}{t_{\rm het}}\\
E_{\rm sup}&=&\frac{1}{\kappa_0\kappa}\left[\Delta P_{\rm sup}-\frac{t_{\rm inf}+t_{\rm sup}}{t_{\rm het}}\Delta\bar{P}\right]+\frac{\Delta V}{t_{\rm het}}\,.
\end{eqnarray}
\end{subequations}

The above equations hold as long as the QD is empty -- which is also often desired experimentally. Neglecting quantum confinement in a first approximation, the QD is empty as long as the conduction band edge is above the Fermi energy, and the valence band edge below the Fermi energy throughout the dot. Assuming $E_{\rm QD}<0$ (which is the case here), the QD is therefore free from holes if $E_v({\rm GaN})-eV(z=-t_{\rm sup}-t_{\rm QD})<\mu$, and free from electrons if $E_c({\rm GaN})-eV(z=-t_{\rm sup})>\mu$. Using Eqs. (\ref{eqmu}), (\ref{eqDV}) and (\ref{eqEinfsup}), these conditions respectively translate into the following constraints on $t_{\rm inf}$ and $t_{\rm sup}$: 
\begin{subequations}
\label{eqempty}
\begin{eqnarray}
t_{\rm inf}&<&\frac{E_{\rm g}({\rm GaN})}{eE_{\rm inf}} \\
t_{\rm sup}&<&\frac{\Delta V}{E_{\rm sup}} \,.
\end{eqnarray}
\end{subequations}
Assuming fixed pyro- and piezoelectric polarizations, $E_{\rm inf}$ and $E_{\rm sup}$ are independent on $t_{\rm inf}$ and $t_{\rm sup}$ for given $t_{\rm QD}$ and $t_{\rm het}$. Equations (\ref{eqempty}) then show that the QD can be empty only in a finite range of positions within the heterostructure. The QD is indeed filled with electrons if it is too far from the surface, and filled with holes if it is too far from the pillar. Note, however, that quantum confinement will practically hinder the charging of the QDs by rising the electron and hole energies. The above constraints thus provide safe bounds for the design of nanowire heterostructures.

Equations (\ref{EqEQD}) and (\ref{eqEinfsup}) also give an estimate of the charge densities $\sigma_s$ and $\sigma_p$ accumulated at top surface and interface with the pillar, respectively. The continuity equation for the electric field indeed reads at this interface:
\begin{equation}
\kappa_0\kappa\left(E_{\rm inf}-E_{\rm pil}\right)=\sigma_p\,,
\end{equation}
where $E_{\rm pil}$ is the electric field in the pillar. Since $E_{\rm pil}$ decreases rapidly away from the interface,
\begin{equation}
\sigma_p\simeq\kappa_0\kappa E_{\rm inf}\,.
\end{equation}
Assuming that the tip of the nanowire is charge neutral at equilibrium, we then get:
\begin{equation}
\sigma_s\simeq-\left(\sigma_p-\Delta P_{\rm inf}+\Delta P_{\rm sup}\right)\,.
\end{equation}
We can further split $\sigma_s$ and $\sigma_p$ into polarization and induced charges:
\begin{subequations}
\begin{eqnarray}
\sigma_p&=&\Delta P_{\rm pil}-n_p e \\
\sigma_s&=&P_{\rm sup}+n_s e \,,
\end{eqnarray}
\end{subequations}
where $\Delta P_{\rm pil}=P_{\rm pil}-P_{\rm inf}$, $n_p$ is the density of the electron gas at the interface with the pillar, and $n_s$ is the density of charges (ionized surface states+holes) at the top surface. The latter thus finally read:
\begin{subequations}
\label{eqnsnp}
\begin{eqnarray}
n_p e&\simeq&P_{\rm pil}-P_{\rm QD}+\frac{t_{\rm inf}+t_{\rm sup}}{t_{\rm het}}\Delta\bar{P}-\kappa_0\kappa\frac{\Delta V}{t_{\rm het}} \\
n_s e&\simeq&-P_{\rm QD}+\frac{t_{\rm inf}+t_{\rm sup}}{t_{\rm het}}\Delta\bar{P}-\kappa_0\kappa\frac{\Delta V}{t_{\rm het}}\,. \label{eqns}
\end{eqnarray}
\end{subequations}
Note that $t_{\rm het}$ must be large enough for the electron gas to form at the interface with the pillar ($n_p>0$),\cite{Ibbetson00} but this is usually not limiting the design of the heterostructure.

We can get a rough estimate of $E_{\rm QD}$, $n_p$ and $n_s$ by neglecting piezoelectricity [$P_{\rm inf}=P_{\rm sup}=P_0({\rm AlN})$, $P_{\rm pil}=P_{\rm QD}=P_0({\rm GaN})$]. We then get from equations (\ref{EqEQD}) and (\ref{eqns}):
\begin{subequations}
\begin{eqnarray}
E_{\rm QD}&=&-\frac{1}{\kappa_0\kappa}\frac{t_{\rm inf}+t_{\rm sup}}{t_{\rm het}}\left[P_0({\rm GaN})-P_0({\rm AlN})\right]+\frac{\Delta V}{t_{\rm het}} \label{eqEQD2} \\
n_p e&=&\frac{t_{\rm inf}+t_{\rm sup}}{t_{\rm het}}\left[P_0({\rm GaN})-P_0({\rm AlN})\right]-\kappa_0\kappa\frac{\Delta V}{t_{\rm het}} \\
n_s e&=&-\frac{t_{\rm QD}P_0({\rm GaN})+\left(t_{\rm inf}+t_{\rm sup}\right)P_0({\rm AlN})}{t_{\rm het}}-\kappa_0\kappa\frac{\Delta V}{t_{\rm het}}\,.
\end{eqnarray}
\end{subequations}
As a simple example, the pyroelectric field in a 4 nm thick QD embedded in an infinitely long nanowire ($t_{\rm het}\to\infty$) would be $|E_{\rm QD}|=7.03$ MV/cm. In a finite heterostructure with $t_{\rm het}=20$ nm, the induced charges screen this field down to $|E_{\rm QD}|=3.47$ MV/cm. The density of the electron gas at the interface with the GaN pillar is then $n_p=1.73\times10^{13}$ cm$^{-2}$, while the total density of charges (surface states+holes) at the top surface is $n_s=3.85\times 10^{13}$ cm$^{-2}$. Therefore, the Fermi level is actually pinned in the valence band of AlN as long as the total (donor) surface states density is lower than $N_{\rm crit}^+=3.85\times 10^{13}$ cm$^{-2}$. This critical density, although large, is yet not unreasonable for bare nanowire surfaces. We will however give further evidence in paragraph \ref{sectionelectronic} that the Fermi level is pinned at (or at least close to) the valence band edge of AlN.

\begin{figure}
\includegraphics[width=.66\columnwidth]{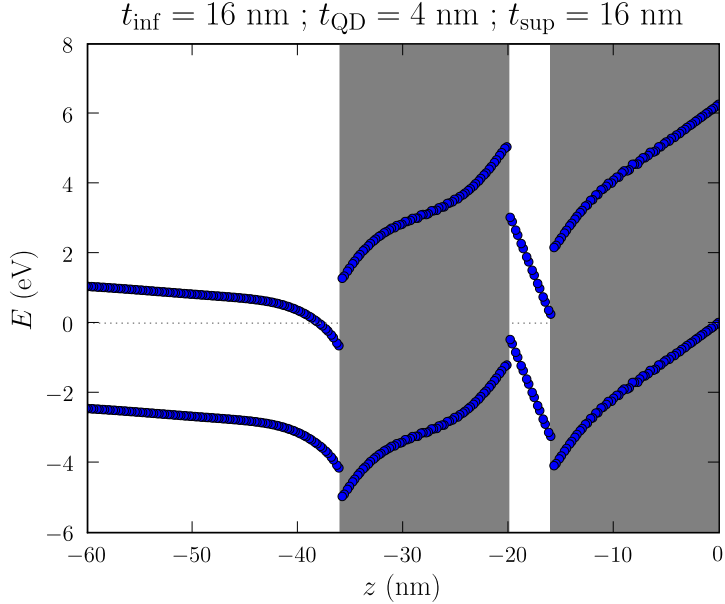}
\caption{(Color online) The conduction band edge energy $\varepsilon_c(\vec{r})=E_c-eV(\vec{r})$ and the valence band edge energy $\varepsilon_v(\vec{r})=E_v-eV(\vec{r})$ along the axis of a nanowire with $t_{\rm QD}=4$ nm and $t_{\rm inf}=t_{\rm sup}=16$ nm. The reference of energy is the chemical potential $\mu=0$. The position of the AlN barriers is outlined in gray. The electric field is not constant in the barriers due to the inhomogeneous strains and piezoelectricity.}
\label{figbands_16_16}
\end{figure}

We have tested this simple 1D model against the numerical solution of equations (\ref{eqnp})--(\ref{eqv}). It gives excellent account of the electric field in the QD when $t_{\rm het}\lesssim 2R$. The effects of the non-intentional doping are indeed negligible with respect to the amount of charges transferred by the pyro- and piezoelectric field. This model however tends to overestimate $n_p$ (as the electric field in the pillar actually decreases over tens of nanometers) and thus overestimates $n_s$ (by around 25\% in the above example). Also, the piezoelectric polarization and field become inhomogeneous in thick heterostructures, as the strains are maximum at the interfaces and relax in between\cite{Niquet06b,Niquet08} (see Fig. \ref{figbands_16_16}). The 1D model above is nonetheless very helpful in understanding trends and guiding the design of nanowire heterostructures.

\subsection{Discussion}

\begin{figure}
\includegraphics[width=.66\columnwidth]{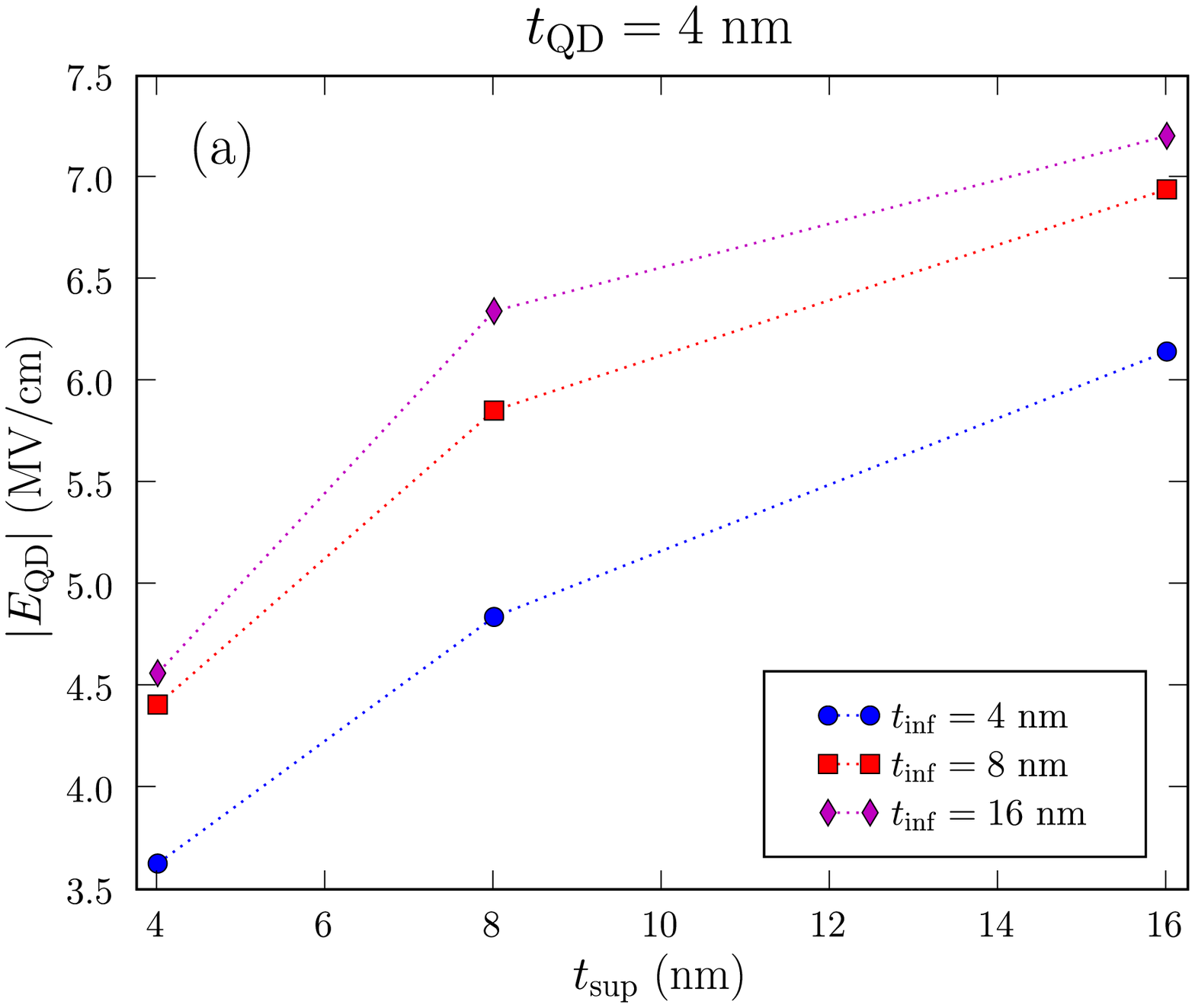} \\
\includegraphics[width=.66\columnwidth]{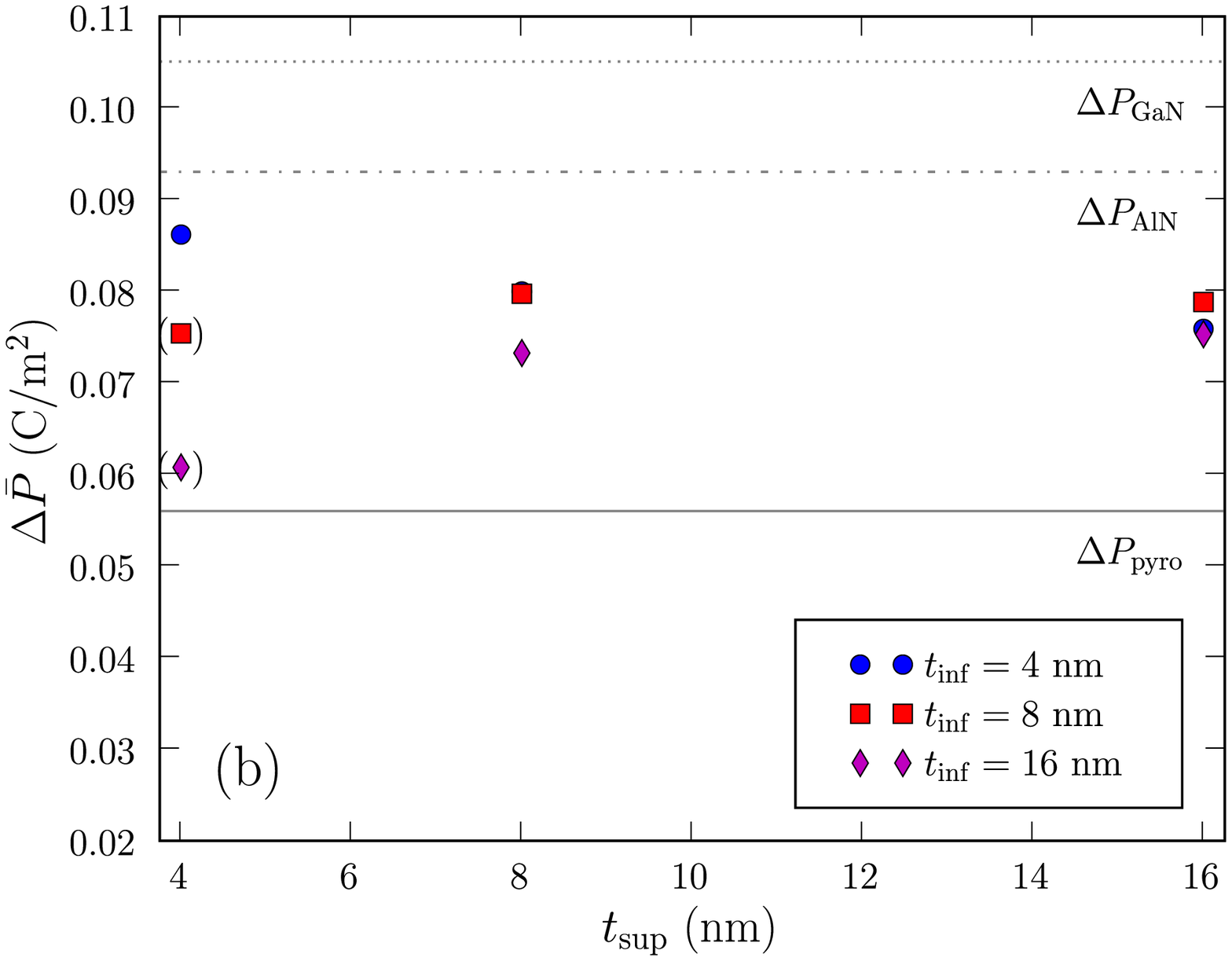}
\caption{(Color online) (a) The amplitude of the electric field $|E_{\rm QD}|$ in the GaN QD as a function of $t_{\rm inf}$ and $t_{\rm sup}$ ($t_{\rm QD}=4$ nm). (b) The average polarization discontinuity $\Delta\bar{P}$ deduced from (a) and Eq. (\ref{EqEQD}). The two dots in parenthesis are charged with holes.}
\label{figE}
\end{figure}

The amplitude of the electric field $|E_{\rm QD}|$, computed with Eqs. (\ref{eqnp})--(\ref{eqv}) as the difference of potential along the QD axis divided by $t_{\rm QD}$, is plotted in Fig. \ref{figE}a as a function of $t_{\rm inf}$ and $t_{\rm sup}$ ($t_{\rm QD}=4$ nm). As expected from Eq. (\ref{eqEQD2}), the electric field increases with the total thickness $t_{\rm het}$ of the heterostructure, and ranges from $\simeq 3.5$ MV/cm for $t_{\rm het}\simeq12$ nm to $>7$ MV/cm for $t_{\rm het}=36$ nm. The electric field is slightly higher than expected from the spontaneous polarization, and does not fulfill the symmetry relation $E_{\rm QD}(t_{\rm inf},t_{\rm sup})=E_{\rm QD}(t_{\rm sup},t_{\rm inf})$ due to piezoelectricity. This is further emphasized in Fig. \ref{figE}b, which represents the average $\Delta\bar{P}$ obtained by inverting Eq. (\ref{EqEQD}) with the data of Fig. \ref{figE}a. Three horizontal lines are also plotted on this figure for reference. $\Delta P_{\rm pyro}=0.056$ C/m$^2$ is the spontaneous polarization discontinuity at the GaN/AlN interface, which should provide a lower bound for $\Delta\bar{P}$. $\Delta P_{\rm GaN}=0.105$ C/m$^2$ is the spontaneous and piezoelectric polarization discontinuity in a heterostructure biaxially strained onto GaN, and $\Delta P_{\rm AlN}=0.093$ C/m$^2$ is the polarization discontinuity in a heterostructure biaxially strained onto AlN, which are the expected limits for thin and thick barriers, respectively. The actual $\Delta\bar{P}$ lies between these bounds, as an evidence for piezoelectricity. The piezoelectric field, though still significant, is lower than in a 2D AlN/GaN/AlN quantum well, due to strain relaxation. The variations of $\Delta\bar{P}$ result from a complex interplay between strain relaxation and charging (see discussion below). It is nonetheless worthwhile to note that a very good approximation to the electric field can be obtained with a constant $\Delta\bar{P}\simeq0.077$ C/m$^2$ (for given $t_{\rm QD}$ and $R$) in a wide range of $t_{\rm inf}$ and $t_{\rm sup}$. The value of $\Delta\bar{P}$ slightly increases with decreasing $t_{\rm QD}$, up to $\Delta\bar{P}\simeq0.082$ C/m$^2$ for $t_{\rm QD}=1$ nm.

As discussed above, the QDs might not be empty if they are are too far from the surface or from the pillar. Eqs. (\ref{eqnp})--(\ref{eqv}) do not, however, properly take quantum confinement into account. We have therefore refined the assessment of the charge state of the QDs with the tight-binding model: we have tentatively assumed that the QDs were empty (setting $N_{\rm v}=N_{\rm c}=0$ in the dots so that they are free of carriers), and checked the position of the tight-binding band edges with respect to the Fermi energy. We find that all the QDs of Fig. \ref{figE} are actually empty, except those with $t_{\rm sup}=4$ nm and $t_{\rm inf}\ge8$ nm, which are filled with holes. As expected from Eq. (\ref{eqns}), the total charge density in the AlN barriers increases with $t_{\rm het}$, from $n_s=2.39\times 10^{13}$ cm$^{-2}$ for $t_{\rm inf}=t_{\rm sup}=4$ nm, to $n_s=3.54\times 10^{13}$ cm$^{-2}$ for $t_{\rm inf}=t_{\rm sup}=8$ nm, and $n_s=4.25\times 10^{13}$ cm$^{-2}$ for $t_{\rm inf}=t_{\rm sup}=16$ nm. The Fermi energy is therefore pinned in the valence band of AlN at the top surface in all heterostructures considered here ($n_s>N_{\rm tot}^+=2\times 10^{13}$ cm$^{-2}$).

We would finally like to discuss the role of the lateral surface states. The top surface states play a key role by releasing electrons in the GaN pillar, thereby screening the pyro- and piezo-electric field. The occupied lateral surface states of the upper and lower AlN barrier also act as a (secondary) source of electrons. Most of these extra electrons (as well as the donor electrons) are, however, trapped by the empty lateral surface states of the GaN pillar. As a consequence, the GaN pillar is effectively depleted far away from the heterostructure, and the Fermi level is pinned $\simeq 1.25$ eV below the conduction band edge (see Fig. \ref{figbands_8_8}).\cite{Simpkins08} This does not, however, have significant influence on the physics of the heterostructure. 


\section{Electronic properties of GaN/AlN nanowire heterostructures}
\label{sectionelectronic}

We now discuss the electronic and optical properties of the GaN/AlN nanowire heterostructures, and compare our results with experimental data.

\begin{figure}
\includegraphics[width=.66\columnwidth]{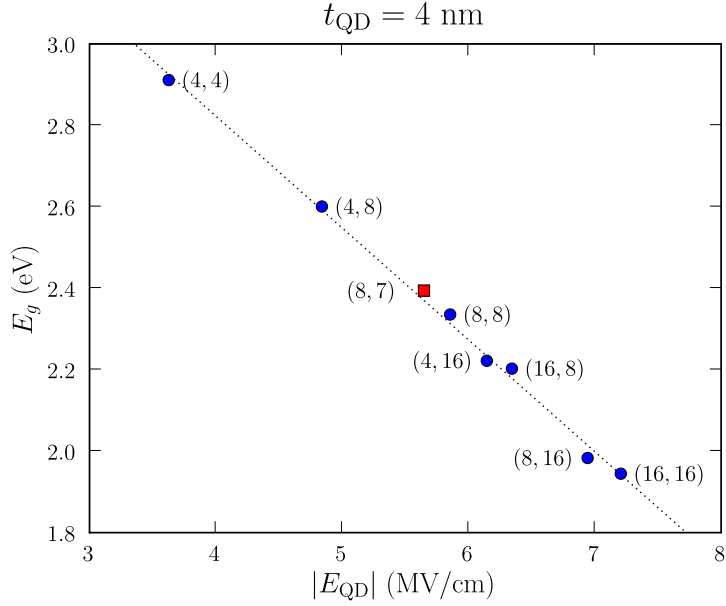}
\caption{(Color online) The band gap energy of GaN QDs as a function of the electric field ($t_{\rm QD}=4$ nm). The dotted line is a guide to the eye. The corresponding ($t_{\rm inf}$, $t_{\rm sup}$) are given (in nm) between parenthesis. The red square is the experimental structure discussed in the text.}
\label{figEg}
\end{figure}

\begin{figure}
\includegraphics[width=.66\columnwidth]{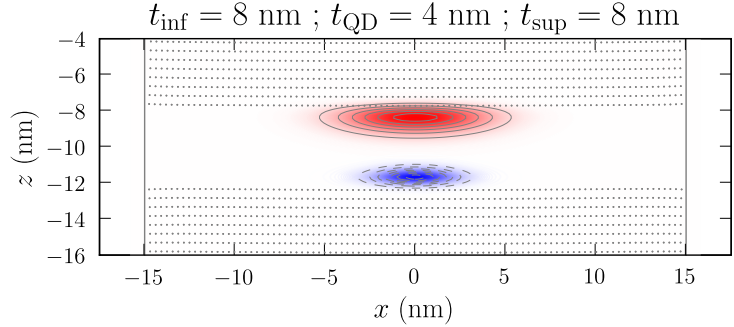}
\caption{(Color online) The lowest hole (red/solid contour lines) and electron (blue/dashed contour lines) wave functions in a GaN QD ($t_{\rm QD}=4$ nm ; $t_{\rm inf}=t_{\rm sup}=8$ nm). The gray dots are the Al atoms in the AlN barriers.}
\label{figwfn}
\end{figure}

The tight-binding band gap energy $E_g$ of empty 4 nm thick GaN QDs is plotted as a function of the electric field $E_{\rm QD}$ in Fig. \ref{figEg}. The corresponding values of $t_{\rm inf}$ and $t_{\rm sup}$ are reported between parenthesis. The excitonic correction is not included in this calculation and should further decrease the optical band gap by at most $\simeq25$ meV. The band gap energy is strongly red-shifted (below the bulk value) by the electric field (Stark effect). It depends almost linearly on $E_{\rm QD}$ and spans around 1 eV in the investigated range of $t_{\rm inf}$ and $t_{\rm sup}$. The ground-state electron and hole wave functions of a particular QD ($t_{\rm inf}=t_{\rm sup}=8$ nm) are plotted in Fig. \ref{figwfn}. As expected, the electron is confined at the upper interface, while the hole is confined at the lower interface of the QD. The electron and hole are, interestingly, both localized around the axis of the nanowire, where the strains are maximum, hence the piezoelectric field slightly larger than at the surface. This helps preventing one of the carriers from being trapped by the charged lateral surface states.\cite{Simpkins08}

\begin{figure}
\includegraphics[width=.66\columnwidth]{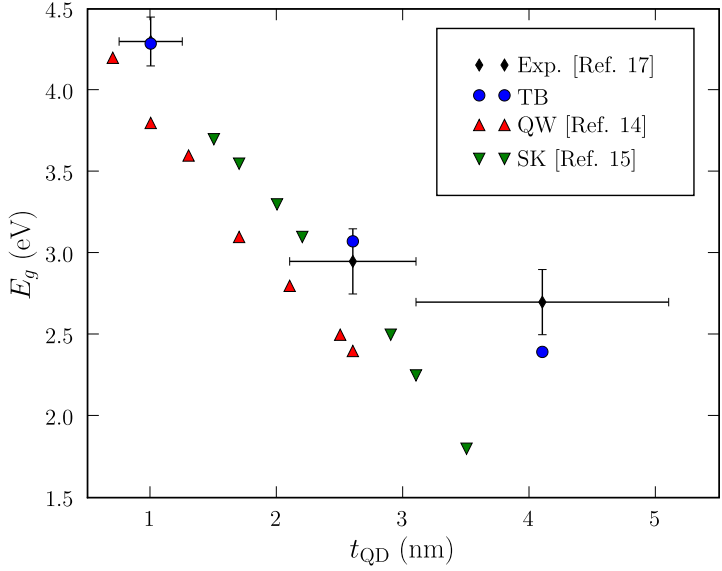}
\caption{(Color online) The calculated (TB) and experimental\cite{Renard09} (Exp.) band gap energies of the GaN QDs. Experimental GaN/AlN quantum wells\cite{Adelmann03} (QW) and Stranski-Krastanov (SK) quantum dots\cite{Bretagnon06} with similar sizes are also reported for comparison.}
\label{figexp}
\end{figure}

Finally, we compare our theoretical predictions with the experimental results of Ref. \onlinecite{Renard09}. In this work, the room-temperature luminescence of 1 to 4 nm thick GaN QDs embedded in 30 nm diameter nanowires showed clear evidence of the quantum confined Stark effect. We have therefore computed the electronic structure of a 1 nm thick QD ($t_{\rm inf}=10$ nm, $t_{\rm sup}=8$ nm), of a 2.5 nm thick QD ($t_{\rm inf}=9$ nm, $t_{\rm sup}=10$ nm) and of a 4 nm thick QD ($t_{\rm inf}=8$ nm, $t_{\rm sup}=7$ nm). The geometry and thickness of the barriers were chosen after a detailed analysis of the experimental TEM images.\cite{Renard09b} The calculated and experimental band gap energies are plotted in Fig. \ref{figexp}. All the dots are empty, which is consistent with the observation of the biexciton in the 1 nm thick QDs.\cite{Renard08} The electric field ranges from $5.6$ MV/cm for $t_{\rm QD}=4$ nm to $7.3$ MV/cm for $t_{\rm QD}=1$ nm. It is, as expected, much smaller than in GaN/AlN quantum wells\cite{Adelmann03} (QWs) and Stranski-Krastanov\cite{Bretagnon06} dots (SKs) with similar sizes due to to strain relaxation and screening by the electron gas and surface charges (see the comparison between QWs, SKs and QDs in Fig. \ref{figexp}).

The calculation reproduces the luminescence energies of the 1 and 2.5 nm thick QDs within the error bars, and the downward trend of the electric field with increasing QD size evidenced in the experiment. 
Still, the calculation underestimates the luminescence energy of the 4 nm thick QDs, where the Stark effect is the largest, by about $0.3$ eV. Looking at Fig. \ref{figEg}, this suggests that the electric field in this QD is overestimated by $\simeq 20$ \%. The reason for this discrepancy is unclear at present. Increasing the density of surface states to pin the Fermi level in the bandgap of AlN increases the electric field and ultimately charges the dots. The experimental data have, moreover, been collected at low enough excitation power to prevent screening by photogenerated multiple electron-hole pairs.\cite{Wu09,Ranjan03,Renard09b} The calculated electric field might be affected by the electromechanical coupling\cite{Jogai03b,Lassen08} (influence of the electric field on strains), by the uncertainties in the pyro- and piezoelectric constants of GaN and especially AlN, and by their dependence on strains (non-linear piezoelectricity).\cite{Bester06,Niquet07a} A simple 1D model however shows that the electromechanical coupling should reduce the electric field by at most $\simeq5\%$ (see Appendix \ref{appendix}). Although the nanowires of Ref. \onlinecite{Renard09} are likely metal-polar, their polarity has not actually been assessed experimentally.\cite{Renard09b} Interestingly, we would like to point out that a N-face polarity would give a slightly better agreement between theory and experiment. In that case, the pyro- and piezo-electric field are reversed, so that a hole gas forms at the interface between the GaN pillar and heterostructure and electrons accumulate at the top AlN surface. The difference of potential across the heterostructure [Eq. (\ref{eqDV})] then becomes $\Delta V=E_v({\rm GaN})-E_c({\rm AlN})=-5.45$ eV, which is slightly higher in magnitude than for the metal-face polarity [$\Delta V=E_c({\rm GaN})-E_v({\rm AlN})=4.3$ eV]. As a consequence, the pyro- and piezoelectric field in the QD are better screened, so that the luminescence in the 4 nm thick QDs is raised by $\simeq 150$ meV. Further experiments (for different barrier and QD thicknesses), as well as detailed polarity measurements might therefore be needed to get a complete picture.

\section{Conclusion}

We have modeled the quantum confined Stark effect in $[0001]$-oriented AlN/GaN/AlN nanowire heterostructures using a tight-binding approach. We have taken strain relaxation and band bending into account in the calculation of the pyro- and piezoelectric field. We have shown that strain relaxation reduces the piezoelectric polarization, and that the electric field pulls out electrons from the occupied states of the top surface. These electrons accumulate in the GaN pillar below the heterostructure, thereby screening the pyro- and piezoelectric field. We suggest that the electric field is likely strong enough to pin the Fermi level in or close to the valence band edge of AlN at the top surface. As a consequence, the electric field is significantly reduced with respect to GaN/AlN quantum wells or Stranski-Krastanov quantum dots. This is in agreement with recent experimental data on GaN/AlN nanowire heterostructures.\cite{Renard09} The calculation however overestimates the electric field in thick quantum dots, which calls for further experiments with different geometries and detailed polarity measurements. We have, for this purpose, provided a simple 1D model for the electric field to help the design of such heterostructures. We thank B. Daudin, B. Gayral, J. Renard and C. Bougerol for fruitful discussions.

\appendix

\section{Electromechanical coupling}
\label{appendix}

In this appendix, we give an estimate of the electromechanical coupling and of its effect on the optical band gap of the nanowire heterostructures.

We assume that the heterostructures are biaxially strained onto GaN. According to Fig. \ref{figE}b, this should provide an upper bound to the electric field, hence to the electromechanical coupling. The in-plane strains $\varepsilon_{xx}=\varepsilon_{yy}=\varepsilon_{\parallel}$ in the GaN QD and AlN barrier are then:
\begin{subequations}
\begin{eqnarray}
\varepsilon_{\parallel}&=&0 {\rm \ in\ GaN} \\
\varepsilon_{\parallel}&=&2.47\% {\rm \ in\ AlN.} 
\end{eqnarray}
\end{subequations}
Following Ref. \onlinecite{Jogai03b}, the vertical strain $\varepsilon_{zz}$ reads in each material:
\begin{equation}
\varepsilon_{zz}=-2\frac{c_{13}}{c_{33}}\varepsilon_{\parallel}+\frac{e_{33}}{c_{33}}E_z\,
\label{emc}
\end{equation}
where $c_{13}$ and $c_{33}$ are the macroscopic elastic constants.\cite{Vurgaftman03} The first term is the ``uncoupled'' elastic strain while the second one describes the feedback of the electric field $E_z$ on the structure (electromechanical coupling). Solving Eq. (\ref{emc}) with Eqs. (\ref{EqEQD})--(\ref{eqEinfsup}) for the electric field yields:
\begin{subequations}
\begin{eqnarray}
E_{\rm QD}&=&-8.15 {\rm\ MV/cm\ without\ electromechanical\ coupling} \\
E_{\rm QD}&=&-7.70 {\rm\ MV/cm\ with\ electromechanical\ coupling}
\end{eqnarray}
\end{subequations}
in a 4 nm thick QD with $t_{\rm inf}=8$ nm and $t_{\rm sup}=7$ nm (experimental geometry). The electromechanical coupling can therefore reduce the electric field by at most 5.75\%. According to Fig. \ref{figEg}, a $0.45$ MV/cm decrease of the electric field would account for a $\simeq120$ meV increase of the bang gap. Additionally, the strains in the GaN layer are:
\begin{subequations}
\begin{eqnarray}
\varepsilon_{zz}&=&0 {\rm\ without\ electromechanical\ coupling} \\
\varepsilon_{zz}&=&-0.17\% {\rm\ with\ electromechanical\ coupling.}
\end{eqnarray}
\end{subequations}
Using the interband deformation potential $a_z=-11.3$ eV in GaN,\cite{Vurgaftman03} the strains in the coupled system would further increase the band gap by $\simeq 19$ meV. As a whole, the electromechanical coupling can not, therefore, be expected to increase the band gap by more than $\simeq 140$ meV. The same order of magnitude is obtained assuming the heterostructure is biaxially strained onto AlN.


\begin{thebibliography}{99}
\bibitem{notelong} The nanowires, though shorter than in the experiments of Ref. \onlinecite{Renard09}, are long enough to converge strains and potentials.
\bibitem{noteeij} The strains are defined from the deformations of tetrahedra built from atoms belonging to the same sublattice.
\bibitem{noteV} The heterostructure is placed at the center of the finite difference box, with the nanowire protruding out on one side of the box. Von Neumann (zero normal electric field) boundary conditions are applied on that side of the box, while Dirichlet ($V=0$) boundary conditions are applied on the others. The convergence of the electric field in the GaN QD with respect to the size of the box has been carefully checked.
\end{thebibliography}
\end{document}